\documentclass[prl,twocolumn,showpacs]{revtex4}
\usepackage{graphicx}
\usepackage[usenames]{color}

\begin{document}

\title{Electromagnetic radiation emanating from the molecular nanomagnet $\mathrm{%
Fe_8}$}
\author{Oren Shafir and Amit Keren}
\date{\today}

\begin{abstract}
Photons emitted by transition between the discrete levels of single
molecular magnets might obay the elementary condition for Dicke's
super-radiance. We investigate this possibility in the $\mathrm{{Fe_{8}}}$
molecule where magnetization jumps are known to occur at discrete magnetic
field values. We found energy bursts each time the molecule undergoes a
magnetization jump, confirming their quantum nature. A series of tests
indicated that photons carry out the energy, and that indeed these photons
obey the elementary conditions for super-radiance.
\end{abstract}

\maketitle

\affiliation{Department of Physics, Technion - Israel Institute of Technology, Haifa
32000, Israel}

In recent years, the interest in single molecule magnets (SMM) has grown
widely, mostly because of their quantum tunneling of the magnetization
(QTM)~ \cite{Nanomagnets}. Some of the future potential applications of SMM
are in quantum computation~\cite{Tejada01}~\cite{Leuenberger}, as multi-bit magnetic memory~
\cite{Shafir}, as an essential part in spintronics~\cite{Bogani}, and as an
MRI contrast~\cite{Cage}. More recently the interaction between SMM and
radiation was investigated. Experiments using external micro-wave sources
have been carried out on Fe$_{8}$ in which the absorption and its relation
to the magnetization curve were studied \cite{Bal, Sorace, Petukhov05,
Bal06, Petukhov07, Cagea}.

In addition, it was proposed in theoretical works that single-molecule
magnets could be used to generate Dicke's super-radiance (SR) \cite
{Chudnovsky02, Yukalov02, Henner, Chudnovsky04, Yukalov05, Tokman}. In this
radiative process, a short intense pulse of light emanates from a molecular
system due to interactions via the electro-magnetic field. For
super-radiance the photon wave length must be similar to the sample size 
\cite{Dicke}. Following these works, Tejada \emph{et al.} reported that
during magnetization avalanches of the molecular magnet $\mathrm{{Mn_{12}}}$%
, radiation was released~\cite{Tejada04, Hernandez}. In the same year, Bal 
\emph{et al.} were also looking for this phenomenon, but with the additional
possibility of being able to analyze the radiation frequency~\cite{Bal04}.
However, they could only place an experimental upper bound on SR emission
from $\mathrm{{Mn_{12}}}$. As far as we know, no attempt has been made to
measure the energy bursts from $\mathrm{{Fe_{8}}}$ molecule.

Here we report the experimental detection of radiation emission from $%
\mathrm{{Fe_{8}.}}$ These molecules have spin $S=10$ and high magnetic
anisotropy that corresponds to a ~27~K energy barrier between spin
projection $S_{z}=\pm 10$ and $S_{z}=0$, in zero external field. These
molecules show QTM at regularly spaced steps in the hysteresis loop~\cite
{Wernsdorfer}.

\begin{figure}[tbp]
\includegraphics[width=3.2093in]{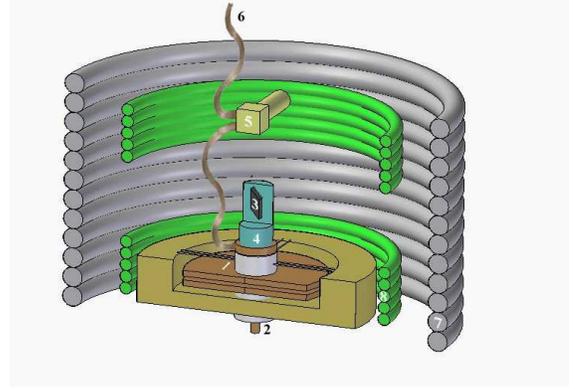} {}
\caption{(Color online) Cross sectional view of the Faraday balance with: (1) movable plate
of the capacitor, (2) screw for capacitor's fixed plate height adjustment,
(3) sample, (4) PCTFE, (5) gold plated casing of the thermometer, (6)
thermal link to the DR mixing chamber, (7) main coil, (8) gradient coils. }
\label{fig1}
\end{figure}

The magnetization is measured using a Faraday force magnetometer as depicted
in Fig.~\ref{fig1}. The design of the magnetometer was dictated by a
different experiment concerning H nuclear magnetic resonance during field
sweep; this experiment will be presented elsewhere. The main objective in
the design was to avoid having any metallic parts next to the sample. The
phenomenon described here was discovered by accident. The Faraday force
magnetometer is mounted in the inner vacuum chamber of a dilution
refrigerator (DR), equipped with a main superconducting magnet that produces
the field $H$, and two oppositely wound superconducting magnets that produce
a field gradient.

The sample is grown by the method described in Ref.~\cite{Weighardt} and is
20 $\mathrm{mm}^{3}$. It is oriented with its easy axis parallel to the
magnetic field $H$ and mounted on the small load-sensing device. The device
is made of two parallel plates variable capacitor. The movable plate is
suspended by two pairs of orthogonal crossed 0.2~mm diameter phosphor bronze
wires attached to it with epoxy. The static plate was mounted on an epoxy
screw, for adjusting the initial capacity $C_{0}$. When the sample is
subjected to a spatially varying magnetic field $B$, it will experience a
force ${\mathbf{F}}=M_{z}(\partial B_{z}/\partial z)\hat{z}$. This force is
balanced by the wires. The displacement of the plate is proportional to $%
\mathbf{F}$ and can be detected as a capacitance $C$ change. The total
capacitance response is then given by 
\[
\frac{1}{C_{0}}-\frac{1}{C}=aM_{z}\frac{\partial B_{z}}{\partial z} 
\]
where $a$ is a constant that depends on the elastic properties of the wires.
This design is discussed further in Sakakibara et al~\cite{Sakakibara}.

The sample is glued to Poly-Chloro-TriFluoro-Ethylene (PCTFE), a
fluorocarbon-based polymer, which has no H atoms and is suitable for
cryogenic applications. The bottom of the PCTFE is connected by a thermal
link to the DR mixing chamber which produces the cooling, and to the movable
plate. Approximately $2$~cm above the sample, on the thermal link, there is
a calibrated thermometer (RuO$_{2}$ R2200) in a gold plated casing. It is
important to mention that the sample is in vacuum with no exchange gas, and
therefore its temperature $T$ is not exactly the same as the temperature of
the thermometer. However, this is not a problem in our experiment since
below $400$~mK the magnetization jumps of Fe$_{8}$ are
temperature-independent~\cite{Sangregorio}. Finally, when needed a copper
cover can be added which blocks the line of sight between the sample and the
thermometer.

\begin{figure}[tbp]
\includegraphics[width=3.2093in]{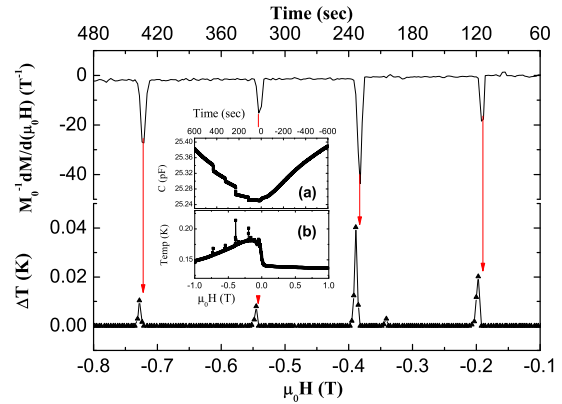} {}
\caption{(Color online) Normalized derivative of the magnetization extracted from the
capacitance (see text) and temperature spikes vs. magnetic field swept from
positive to negative. The changes in the magnetization are followed by an
increase in the temperature, indicating release of energy. Raw date is in
the inset: (a) capacitance (which represents magnetization) and (b)
temperature vs. magnetic field swept from positive to negative. Steps in the
capacitance indicate QTM in the sample. }
\label{fig2}
\end{figure}

In the experiment we apply a field of +1 T and wait until thermal
equilibrium is reached. We then record the capacitance, temperature, and
field value as the field is swept from +1 T to -1 T at a rate of 0.1 T/min.
The capacitance vs. the applied magnetic field (and time) is shown in inset
(a) of Fig.~\ref{fig2}. When the field is positive the capacitance is a
smooth function of the field. This is because the spins are at their ground
state for all positive fields and have nowhere to tunnel to. Once the field
becomes negative, clear jumps in the capacitance are observed, indicating
jumps in the magnetization that are taking place as the magnetization is
tunneling between states. In inset (b) of Fig.~\ref{fig2} we show the
temperature reading of the thermometer. For positive field the temperature
is quite stable. At zero field there is a big and broad increase in the
temperature. This is caused by an eddy
currents developing in the copper wires due to the change in the sweep rate
during the transition from positive to negative field. At negative fields
there is a mild decline in the temperature, accompanied by clear temperature
spikes.

In principle, $C$ should have been constant for $H>0$ since the
magnetization is constant. However, in a DR it is difficult to place the
sample in the center of the main magnet, and the gradient has some field
dependence. The measurements at $H>0$ could be used to calibrate the field
gradient. A simpler approach is to present 
\[
\frac{2}{\Delta C}\frac{dC}{d\mu _{0}H}=\frac{1}{M_{0}}\frac{dM}{d\mu _{0}H} 
\]
where $\Delta C$\ is the difference in capacitance between $H=0$ and $H=1T$,
and $M_{0}$ is the saturation magnetization. This quantity is significant
only at the jumps. We also subtracted from $T$ a polynomial fit to the mild
temperature decline for negative fields. The resulting $(1/M_0 )dM/d\mu _0 H$
and $\Delta T$ are shown in Fig.~\ref{fig2}. It is now clear that the
thermal spikes of a few tens of mili-Kelvin occur about 1 sec after the
capacitance (magnetization) jumps, and that every magnetization jump is
accompanied by a thermal spike. The thermal spikes begin at the lowest field
where tunneling is taking place, indicating that they involve transitions
between the lowest-lying states of the molecular spin. This is a very
different situation from $\mathrm{{Mn_{12}}}$ where the energy bursts are
believed to be due to transitions between high-lying states~\cite{Tejada04}.
Finally, in $\mathrm{{Fe_{8}}}$, the bursts take place in a region where
tunneling is temperature-independent. This should make their analysis much
simpler.

\begin{figure}[tbp]
\includegraphics[width=2.4in]{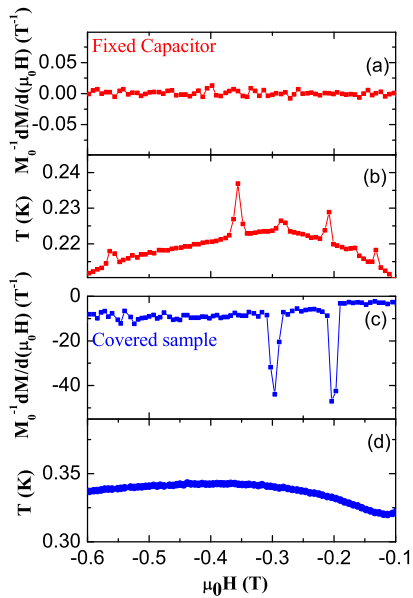}
\caption{(Color online)Test cases: The capacitance (a) and temperature (b) vs. magnetic
field swept from positive to negative with jammed capacitor. The rise in the
temperature indicates the change in the magnetization. The normalized
derivative of the magnetization (c) and temperature (d) vs. magnetic field
(same sweeping direction) with covered sample. The change in the
magnetization is not followed this time by an increase in the temperature. }
\label{fig3}
\end{figure}

\textit{A priori}, there could be many reasons for the thermal spikes. The
first that comes to mind is heating from the moving part of the capacitor.
To disqualify this possibility we jammed the movable capacitor plate by
raising the lower plate until they touch each other, and repeated the
measurement. The results are presented in Fig.~\ref{fig3}(a-b). Because the
capacitors' plates were jammed, there is no change in the capacitance, but
the spikes in the temperature are still present.

Another source of heating could be phonons. Since the entire system is in
vacuum, the energy could reach the thermometer only via the copper wire
thermal link. To check this possibility we performed two experiments. First,
we moved the thermometer to a separate copper wire, thermally linked
directly to the mixing chamber, but not to the sample. We confirmed that the
results presented in Fig.~\ref{fig2} are reproducible in this configuration
(not shown). Second, we blocked the line of sight between sample and
thermometer by covering the sample with a copper cylinder. The results are
depicted in Fig.~\ref{fig3}(c-d). The steps in the capacitance are still
seen, although not all of them and they are somewhat broader for a reason
that is not clear to us. Perhaps the force acting on the sample causes it to
fracture after many field and thermal cycles. In contrast, the jumps in the
temperature disappeared completely. The last two experiments confirmed that
the cause of the temperature spikes is electro-magnetic radiation and not
phonons.

A question that should be asked is how come this phenomenon has not been
seen before. We believe that all the experiments with Fe$_{8}$ used exchange
gas or liquid as a cooler, and not a thermal link. In the former case, the
radiation emitted from the sample is hard to detect. Moreover, most
experiments have been done with small crystals to prevent avalanches, so the
radiation was weak.

\begin{figure}[tbp]
\includegraphics[width=3.2093in]{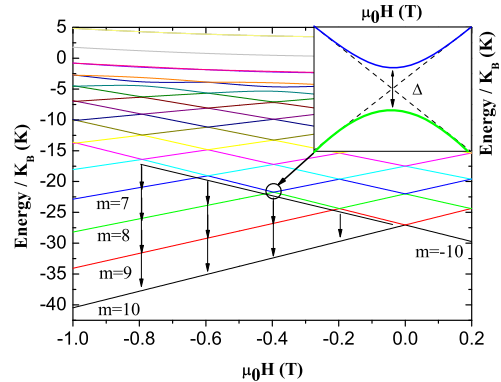}
\caption{(Color online) Zeeman diagram of the 21 levels of the $S=10$ manifold of $\mathrm{{%
Fe_{8}}}$ as a function of the field applied along the easy axis and the
quantum numbers m. The inset is a zoom on the level crossing, which, in
fact, is an avoided crossing with energy split $\Delta $~\protect\cite
{Wernsdorfer}. }
\label{fig4}
\end{figure}

Next we identify the energy levels that participate in the transitions. The
main part of the Hamiltonian is given by 
\begin{equation}
\mathcal{H}=DS_{z}^{2}+E\cdot \left( {S_{x}^{2}-S_{y}^{2}}\right) +g\mu
_{B}H_{z}S_{z}
\end{equation}
where z is the direction of the large uniaxial anisotropy, $S_{x}$, $S_{y}$,
and $S_{z}$ are the three components of the total spin operator, $%
D/k_{B}=-0.292$ K and $E/k_{B}=0.046$ K are the axial and the rhombic
anisotropy parameter, respectively ($k_{B}$ is the Boltzmann factor), $\mu
_{B}$ is the Bohr magneton, and the last term of the Hamiltonian describes
the Zeeman energy associated with an applied field $H$~\cite{Barra, Caciuffo}%
. The energy as a function of field and corresponding level quantum number m
is shown in Fig.~\ref{fig4}~\cite{Wernsdorfer}. In the inset, a zoom view of
the avoided level crossing taking place at $\mu _{0}H=-0.4$ T is presented.
There are two possible transitions. The first possibility, suggested in the
original SR theory, is that the photon is emitted by transition between the
avoided levels as indicated by the vertical arrow in the inset of Fig.~\ref
{fig4}. The photon energy in this case equals that of the tunnel splitting
which is ~$10^{-6}$ K~\cite{Wernsdorfer}. The second possibility is that
photons are emitted due to transition between states with the same sign of
their quantum number \emph{m} as indicated by the solid arrows in the main
panel of Fig.~\ref{fig4}~\cite{Tejada04, Bal04}. In the case of $\mathrm{{%
Mn_{12}}}$ these were high-lying thermally excited states such as $m=1$ to $%
m=2$. In the experiment presented here these must be low-lying states. In
this case the photon energy is ~$\sim 5$~K. The difference in photon energy
expected from these two possibilities is huge and can easily be
distinguished. 

To determine the energy released by the sample we have to convert the size
of the thermal spikes to the energy detected by the resistor. For this
purpose, we measured the energy needed to change the temperature of the
thermometer by the same amount as in Fig.~\ref{fig2}, when the energy is
injected directly into it. The temperature is determined by four-wire
resistance measurement, with very low current of 0.7$~\mu $A. Changing the
current to 10$~\mu $A for ~0.5~sec and immediately after measuring it with
0.7$~\mu $A produced a spike similar to the ones shown in Fig.~\ref{fig2}.
The energy needed to produce these thermal spikes is ~0.25$~\mu J$.

To estimate this energy theoretically we consider the possibility where by
sweeping the field from positive to negative, the tunneling that is taking
place at $\mu _{0}H=-0.4$ T is from $m=-10$ to $m=8$, followed by a
transition from $m=8$ to $m=9$ to $m=10$. Judging from the relative area of
the magnetization derivative peaks in Fig.~\ref{fig2}, about 0.4 of the
total spins tunnel at this crossing. The expected energy release after the
tunneling is twice ~5 K (see Fig.~\ref{fig4}) or $1.4\times 10^{-22}$ J. The 
$20$ mm$^{3}$ sample, with ~$2$~nm$^{3}$ unit cells~\cite{Weighardt}, has $%
10^{19}$ molecules. Therefore, the energy that was released is $0.6$~mJ.
Considering the distance between sample and thermometer and its
cross-section, the solid angle of the thermometer is $0.02\pm 0.004$.
Therefore, the energy that should reach it is 12$~\mu J$. This is much
closer to the estimated value discussed above than energy from avoided
levels photon of $10^{-6}$~K. Therefore, it is clear that photons emitted by
transitions between low lying states, and not avoided levels, are
responsible for the thermal spikes.

Having established the energy carrier and the energy source we examine first
the possibility of black body radiation. The temperature of the sample can
increase after the magnetization steps but not too much since we see the
consecutive step. An upper limit is 5 K where steps are no longer observed.
At this temperature Stephan-Boltzmann law would predict a radiation power
two-three order of magnitude smaller than what is needed to produce our
temperature spikes.

Next we consider the possibility of SR. The most important condition for SR
is $\lambda >l$, where $\lambda $ is the photon wave length, and $l$ the
sample size of $2.7$~mm in our case. $\lambda $ for a $5$~K photon is $3$%
~mm. Therefore, this SR condition is obeyed. The second condition is that
the transition rate will be bigger than any other decoherence rate of the
molecular spins. The transition rate for a single molecule emitting a photon
is~\cite{Tejada04} 
\begin{equation}
\Gamma _{1}=\frac{{2g^{2}\mu _{B}^{2}}}{{3\hbar ^{4}c^{3}}}%
(S-m)(S+m+1)(E_{m}-E_{m+1})^{3}.
\end{equation}%
For the $m=8$ to $m=9$ this gives $\Gamma _{1}=10^{-7}$ sec%
$^{-1}$. In the SR case
the minimal transition rate $\Gamma _{SR}=N\Gamma _{1}$, where $N$ is the
total number of molecules in the $m=8$ (without the thermal factor which
exists in $\mathrm{{Mn_{12}}}$). The maximum transition rate is $\Gamma
_{SR}=N^{4}\Gamma _{1}/4$~\cite{Dicke}. This gives $\Gamma _{SR}>10^{11}$ sec%
$^{-1}$. Since there is no temperature dependence of the tunneling in $%
\mathrm{{Fe_{8}}}$ below 400mK it is believed that the source of dephasing
is nuclear moments, and it is given by $\Gamma _{nuclear}\sim 10^{8}$~sec$%
^{-1}$~\cite{Keren}. Therefore, the second SR condition $\Gamma _{SR}\gg
\Gamma _{nuclear}$ is also obeyed. Thus, it is conceivable that the
transitions between low-lying states in $\mathrm{{Fe_{8}}}$ are accompanied
with SR of photons.

Finally, we consider the possibility of classical magnetic dipole radiation.
It was shown in Ref.~~\cite{Chudnovsky02} that since this radiation is a
collective phenomena that conserves the total spin value, it is equivalent
to SR, provided that the relaxation between levels occurs fast enough.
Eq.~18 in Ref.~~\cite{Chudnovsky02} relates the emitted power $I$ to the
second derivative of the magnetization projection by $I=\frac{2}{3c^{3}}%
\left( \frac{d^{2}m_{z}}{dt^{2}}\right) ^{2}$, which could be approximated
as $\frac{2}{3c^{3}}\frac{\Delta m_{z}^{2}}{\Delta t^{4}}$. Using this
relation, our energy burst for the transition between say $m=8$ to $m=10$
can be viewed as dipole radiating classically for $\sim 10$~nsec. \ This
time is much shorter than $1/\Gamma _{1}$ and closer to $1/\Gamma _{SR}$,
hence the equivalence to SR.

We acknowledge helpful discussions with Aharon Gero, Eugene Chudnovsky and Javier Tejada. The work was funded by
Israel Ministry of science, and by the Nevet program of the RBNI center for
nano-technology.

\end{document}